\documentclass[preprint,pdf]{iucr}              % DO NOT DELETE THIS LINE

\begin{document}                  % DO NOT DELETE THIS LINE

     %-------------------------------------------------------------------------
     % The introductory (header) part of the paper
     %-------------------------------------------------------------------------

     % The title of the paper. Use \shorttitle to indicate an abbreviated title
     % for use in running heads (you will need to uncomment it).

\title{Asymmetric rotations and dimerization driven by normal to modulated phase transition in 4-biphenylcarboxy coupled L-phenylalaninate}
%\shorttitle{Short Title}

     % Authors' names and addresses. Use \cauthor for the main (contact) author.
     % Use \author for all other authors. Use \aff for authors' affiliations.
     % Use lower-case letters in square brackets to link authors to their
     % affiliations; if there is only one affiliation address, remove the [a].

\cauthor[a,b]{Somnath}{Dey}{dey@ifk.rwth-aachen.de}{address if different from \aff}
\author[a]{Supriya}{Sasmal}
\author[a]{Saikat}{Mondal}
\author[a]{Santosh}{Kumar}
\author[a]{Rituparno}{Chowdhury}
\author[a]{Debashrita}{Sarkar}
\author[a]{C. Malla}{Reddy}
\author[b]{Lars}{Peters}
\author[b]{Georg}{Roth}
\author[a]{Debasish}{Haldar}

\aff[a]{Department of Chemical Sciences, Indian Institute of Science Education and Research (IISER) Kolkata, Mohanpur 741246 \country{India}}
\aff[b]{Institute of Crystallography, RWTH Aachen University, Jägerstraße 17-19, 52066 Aachen \country{Germany}}

     % Use \shortauthor to indicate an abbreviated author list for use in
     % running heads (you will need to uncomment it).

%\shortauthor{Soape, Author and Doe}

     % Use \vita if required to give biographical details (for authors of
     % invited review papers only). Uncomment it.

%\vita{Author's biography}

     % Keywords (required for Journal of Synchrotron Radiation only)
     % Use the \keyword macro for each word or phrase, e.g.
     % \keyword{X-ray diffraction}\keyword{muscle}

%\keyword{keyword}

     % PDB and NDB reference codes for structures referenced in the article and
     % deposited with the Protein Data Bank and Nucleic Acids Database (Acta
     % Crystallographica Section D). Repeat for each separate structure e.g
     % \PDBref[dethiobiotin synthetase]{1byi} \NDBref[d(G$_4$CGC$_4$)]{ad0002}

%\PDBref[optional name]{refcode}
%\NDBref[optional name]{refcode}

\maketitle                        % DO NOT DELETE THIS LINE

\begin{synopsis}
Coupled molecule of 4-biphenylcarboxy-(L)-phenylalaninate undergoes
normal to commensurately modulated phase transition at $T_{c}$ $\sim$ 124 K
that is characterized by torsional modulation typical of modulated structures
and superstructures of polyphenyls yet unusual owing to asymmetric and unequal rotations
governed by intramolecular and intermolecular constraints.
The transition is presumably governed by significant suppression of atomic displacement parameters 
correlated to evolution of amplitude of atomic modulation functions. 
\end{synopsis}

\begin{abstract}
Amongst the derivatives of 4-biphenylcarboxylic acid and amino acid esters,
the crystal structure of 4-biphenylcarboxy-(L)-phenylalaninate is unusual
owing to its monoclinic symmetry
within a pseudo-orthorhombic lattice.
The distortion is described by disparate rotational property around the chiral centers
($\varphi_{\mathrm{chiral}}$ $\simeq$ -129 degrees and 58 degrees)
of the two molecules in the asymmetric unit.
Each of these molecules comprise of planar biphenyl moieties
($\varphi_{\mathrm{biphenyl}}$ = 0 degrees).
Using temperature dependent single crystal X-ray diffraction experiments
we show that the compound undergoes a phase transition below $T$ $\sim$ 124 K
that is characterized by a commensurate modulation wave vector,
$\mathbf{q}$ = $\delta(101)$, $\delta$ = $\frac{1}{2}$.
The (3+1) dimensional modulated structure at $T$ = 100 K
suggests that the phase transition drives the biphenyl
moieties towards non coplanar conformations
with significant variation of internal torsion
($\varphi^{\mathrm{max}}_{\mathrm{biphenyl}}$ $\leq$ $20$ degrees).
These intramolecular rotations lead to dimerization of the molecular stacks that are described
predominantly by intermolecular tilts and small variations in
intermolecular distances.
Atypical of modulated structures and superstructures of biphenyl and other polyphenyls,
the rotations of individual molecules are asymmetric ($\Delta$$\varphi_{\mathrm{biphenyl}}$ $\approx$ 5 degrees)
while $\varphi_{\mathrm{biphenyl}}$ of one independent molecule is two to four times larger than the other.
Crystal-chemical analysis and phase relations in superspace
suggest multiple competing factors involving intramolecular steric factors,
intermolecular H--C${\cdot}{\cdot}{\cdot}$C--H contacts and
weak C--H${\cdot}{\cdot}{\cdot}$O hydrogen bonds that govern the
distinctively unequal torsional property of the molecules.
\end{abstract}

     %-------------------------------------------------------------------------
     % The main body of the paper
     %-------------------------------------------------------------------------
     % Now enter the text of the document in multiple \section's, \subsection's
     % and \subsubsection's as required.

\section{Introduction}

Molecular biphenyl has been investigated
extensively for its stability
and conformation in different thermodynamic states.
At ambient conditions, the differences in the conjugation states of
the $\pi$-electrons are governed primarily by
the twist about the central C--C single bond in the order:
40 deg - 45 deg in gas phase,
20 deg - 25 deg in solution
and 0 deg (mutually coplanar) in the solid state in centrosymmetric monoclinic space group
$P2_{1}/a$
\cite{bastiansen_o_1949,suzuki_h_1959,trotter_j_1961,hargreaves_a_1962}.
The planar conformation due to constraints from intermolecular interactions
is energetically unfavourable and steric hindrance between the ortho hydrogen atoms
is compensated for by out of plane dynamic disorder and in plane displacements of those hydrogens away
from each other
\cite{hargreaves_a_1962,casalone_g_1968,charbonneau_gp_1976,charbonneau_gp_1977,busing_wr_1983,lenstra_ath_1994}
A recent study has also suggested the role of intramolecular exchange energy between
single bonded carbon atoms in stabilizing the planar conformation \cite{popelier_pla_2019}.
Absorption and fluorescence studies
showed additional bands in their spectra
at low temperatures ($T$)
\cite{hochstrasser_rm_1973,wakayama_ni_1981}.
$T$-dependent Raman spectroscopy and Brillouin scattering experiments have suggested
two phase transitions at $T_{c1}$ = 42 K and $T_{c2}$ = 17 K respectively
\cite{friedman_ps_1974,bree_a_1977,bree_a_1978,ecolivet_c_1983}.
The phase transition at $T_{c1}$ is continuous and governed by a soft mode associated with the
torsion about the central C--C single bond followed by discontinuous changes at $T_{c2}$.
Inelastic neutron scattering experiments on its
deuterated form confirmed the phase transitions with appearance of additional
satellite reflections
\cite{cailleau_h_1979}.
The modulation wave vector ($\mathbf{q}$) was determined
to be $\mathbf{q}_{I}$ = ${\delta}_{a}\mathbf{a}^{*}\,+\,\frac{1}{2}(1-{\delta}_{b})\mathbf{b}^{*}$
and $\mathbf{q}_{II}$ = $\frac{1}{2}(1-{\delta}_{b})\mathbf{b}^{*}$
at the intermediate and low temperature phases respectively
and were found to vary with temperature, suggesting an incommensurate nature of the
modulation
\cite{cailleau_h_1979}.
The modulated structure of the low temperature phase II was
described within an acentric superspace group symmetry
$Pa(0{\sigma}_{2}0)0$
\cite{de_wolff_pm_1974,stokes_ht_2011,van_smaalen_s_2013}
and found to be essentially associated with small modulation of
translation and rotation ($\omega$) normal to the mean molecular plane,
and a significant torsion ($\varphi$) between the phenyl rings
\cite{baudour_jl_1983,petricek_v_1985,pinheiro_cb_2015,schoenleber_a_2011}.
Theoretical studies have suggested that
competition between intramolecular and intermolecular forces
drives the phase transition towards the incommensurately modulated states
\cite{ishibashi_y_1981,benkert_c_1987a,benkert_c_1987b,parlinski_k_1989}.
The fundamental property of flexibility in conformations have made biphenyl
an excellent candidate to tune in multifaceted properties
in materials.
Torsion between the rings has been demonstrated to regulate conductivity of
single molecule biphenyl-dithiol junctions
\cite{vonlanthen_d_2009,mishchenko_a_2010,buerkle_m_2012,jeong_h_2020},
tune in thermopower as function of twist angle
\cite{buerkle_m_2012},
degeneracy of energy states on substrates
\cite{cranney_m_2007}
and theoretically suggest wide band gap semiconducting properties
of its derivatives
\cite{khatua_r_2020}.
On the other hand, biphenyl derivatives have also been reported to influence
and increase
the efficiency of photophysical properties
\cite{Oniwa_k_2013,wei_j_2016}.
Planar biphenyl molecule in solid state favours
maximum intramolecular conjugation of $\pi$ electrons as well
as increases the probability of interactions
between delocalized electrons that could aid in optimal stacking of molecules.
Coupling reaction mechanism \cite{seechurn_cccj_2012}
was successfully employed to synthesize
4-biphenylcarboxy protected amino acid esters
of L-serine, L-tyrosine, L-alanine, L-leucine and L-phenylalanine
\cite{sasmal_2019_1,sasmal_2019_2}.
In solid state, the compounds crystallize either in acentric orthorhombic space group symmetry
$P2_{1}2_{1}2_{1}$ or the monoclinic subgroup $P2_{1}$
\cite{sasmal_2019_1,sasmal_2019_2}.
Crystal packing in these systems is determined by
${\pi}{\cdot}{\cdot}{\cdot}{\pi}$ stacking between the
biphenyl fragments and linear strong hydrogen bonds between the amino acid ester moieties.

\begin{figure}
\includegraphics{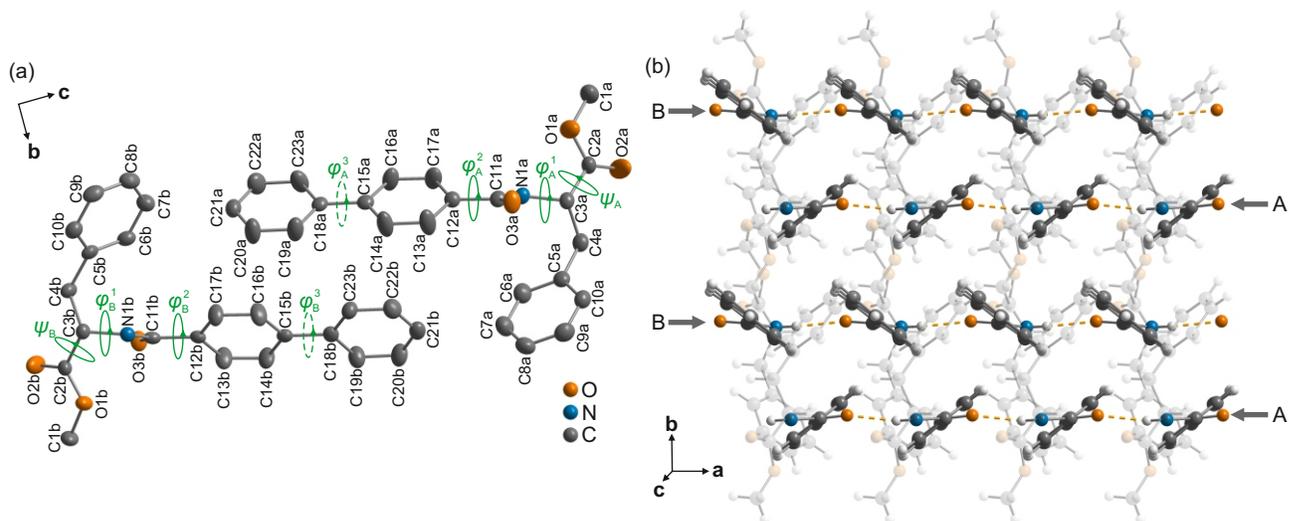}
\caption{(a) Two independent formula units of 4-biphenylcarboxy-(L)-phenylalaninate
(C$_{23}$H$_{21}$NO$_{3}$)
with atomic labels of non-hydrogen atoms (`a' and `b' for molecule `A' and `B' respectively) in phase I at $T$ = 160 K.
${\varphi}_{\mathrm{A}}^{1}$ = -130.1 deg, ${\varphi}_{\mathrm{B}}^{1}$ = 56.1 deg;
${\mid}{\varphi}_{\mathrm{A}}^{2}{\mid}$ = 32.8 deg, ${\mid}{\varphi}_{\mathrm{B}}^{2}{\mid}$ = 30.8 deg;
${\varphi}_{\mathrm{A}}^{3}$ = ${\varphi}_{\mathrm{B}}^{3}$ = 0 deg;
${\psi}_{\mathrm{A}}$ = 36.4 deg, ${\psi}_{\mathrm{B}}$ = 36.2 deg.
Viewing direction along $[\overline{1}00]$.
(b) View of corresponding layered structure along $[111]$ emphasizing biphenyl stacks (AA)$_{\mathrm{n}}$
and (BB)$_{\mathrm{n}}$ along $\mathbf{a}$, (ABAB)$_{\mathrm{n}}$ along $\mathbf{b}$ and
N--H${\cdot}{\cdot}{\cdot}$O bonds (dashed orange) along $[\mp100]$ directions.
Phenyl rings of the ester groups (transparent) stack only along $\mathbf{a}$.
\label{l:biphome_160_labels_packing}}
\end{figure}

We presumed that the biphenyl moieties in these chemically coupled
systems could influence the bioactive amino acid esters and \textit{vice-versa}
with respect to evolution or suppression of translational and rotational degrees of freedom
in their crystal structures at some thermodynamic condition.
Reanalyzing all their crystal structures,
the system of 4-biphenylcarboxy-(L)-phenylalaninate
attracted our attention because the structure appeared to be
similar to the L-tyrosine analogue albeit the monoclinic distortion
(Table~\ref{tab:T_SCXRD},\cite{sasmal_2019_2})
and two crystallographically independent formula units
[$Z^{\prime}$ = 2 \cite{steed_km_2015},
Fig.~\ref{l:biphome_160_labels_packing}(a)]
in the crystal structure of the former.
The torsion about the chiral center
is significantly different for the independent molecules
while the rest of the rotations are similar
[Fig.~\ref{l:biphome_160_labels_packing}(a),
\cite{sasmal_2019_2}].
Each of these molecules consist of coplanar
biphenyl moieties
which are stacked along $\mathbf{a}$
and $\mathbf{b}$ respectively while
the amide groups are connected by intermolecular
N--H${\cdot}{\cdot}{\cdot}$O hydrogen bonds
[Fig.~\ref{l:biphome_160_labels_packing}(b),\cite{sasmal_2019_2}].
In the present study, $T$-dependent phase transition of 4-biphenylcarboxy-(L)-phenylalaninate
has been investigated
using single crystal X-ray diffraction experiments.
The low temperature phase II is found to be a $2a{\times}b{\times}2c$ superstructure
of the high temperature (phase I) structure.
The superstructure is described within the $(3+1)D$ superspace approach
as a commensurately modulated structure
\cite{de_wolff_pm_1974,janner_a_1977,wagner_t_2009,van_smaalen_2012,Janssen_t_2018}.
Structural properties of the phase I and the modulated structure
have been tabulated and compiled within $t$-plots
($t$ = phase of the modulation).
The origin and stability of phase II is discussed in terms
of intra- and intermolecular HC${\cdot}{\cdot}{\cdot}$CH contacts and
intermolecular hydrogen bonds
and plausible mechanism of the
phase transition is suggested.

\section{Experimental}

\subsection{Temperature dependent single crystal X-ray diffraction}

Single crystals of the compound used in this study
were obtained from those reported by
\cite{sasmal_2019_2}.
The crystals were protected in oil under mild refrigeration.
Single crystal X-ray diffraction (SCXRD) experiments were performed
on a Agilent SuperNova, Eos diffractometer employing CuK${\alpha}$ radiation.
Temperature of the crystal was maintained by a open flow nitrogen cryostat
from Oxford Cryosystems.
During cooling, visual inspection of diffraction images
revealed weaker reflections in addition to strong reflections
at low temperatures.
Diffraction images collected at $T$ = 150 K, 140 K and
130 K to 114 K in steps of ${\Delta}T$ = 2 K showed that the
weaker diffuse features appear at $T$ = 124 K that condense into
satellite reflections at $T$ = 122 K
(Table~\ref{tab:T_SCXRD}, Fig. S1 in supporting information).
The transition temperature is significantly higher than
that of molecular biphenyl ($T_{c, \mathrm{biphenyl}}$ = 42 K).
On the other hand,
related polyphenyls
\textit{p}-terphenyl and \textit{p}-quarterphenyl undergo
phase transition towards superstructure phases
at much higher critical temperatures
($T_{c, \mathrm{terphenyl}}$ $\approx$ 190 K \cite{yamamura_y_1998},
$T_{c, \mathrm{quarterphenyl}}$ $\approx$ 233 K \cite{saito_k_1985}).
Complete diffraction data were collected at
$T$ = 160 K and
$T$ = 100 K respectively.

Determination of lattice parameters and data reductions
were performed using the software suite
\textsc{CrysAlisPro}
\cite{CrysAlisPro}
(Table~\ref{tab:T_SCXRD}, Table S1 in supporting information).
Satellite reflections of first order ($m$ = 1) observed below $T_{c}$
could be indexed with modulation wave-vector
$\mathbf{q}$ = (${\sigma}_{1}$,0,${\sigma}_{3}$),
${\sigma}_{1}$ = ${\sigma}_{3}$ $\simeq$ $\frac{1}{2}$
with respect to the basic monoclinic lattice.
Here, the $\mathbf{q}$ = $\frac{1}{2}$(101)
is perpendicular to the $\mathbf{b}$ axis
consistent with monoclinic symmetry
while in molecular biphenyl $\mathbf{q}_{I}$ violates monoclinic symmetry and
$\mathbf{q}_{II}$ is parallel to $\mathbf{b}$
\cite{cailleau_h_1979}.
Using the plugin program \textsc{NADA}
\cite{schoenleber_a_2001}
in \textsc{CrysAlisPro}, deviation of the $\sigma$'s as function of $T$ from a rational value of 0.5
were found to be within their e.s.ds (Table ~\ref{tab:T_SCXRD}),
indicating a commensurate nature of the modulation.
Reflections at $T$ = 100 K were indexed by four integers ($hklm$) using a
basic monoclinic $b$-unique lattice (Table~\ref{tab:T_SCXRD}, Table S1 in supporting information) and
modulation wave vector, $\mathbf{q}$ = ($\frac{1}{2}$,0,$\frac{1}{2}$)
and data integration was performed.
Empirical absorption correction was performed
using \textsc{AbsPack} program embedded in \textsc{CrysAlisPro}.
The monoclinic lattice as well as the reflection conditions suggest
the superspace group $P2_{1}(\sigma_{1}0\sigma_{3})0$
with $\sigma_{1}$ = $\sigma_{3}$ = $\frac{1}{2}$
\cite{stokes_ht_2011,van_smaalen_s_2013}.

\subsubsection{Structure refinement of the modulated structure}

\begin{figure}
\includegraphics{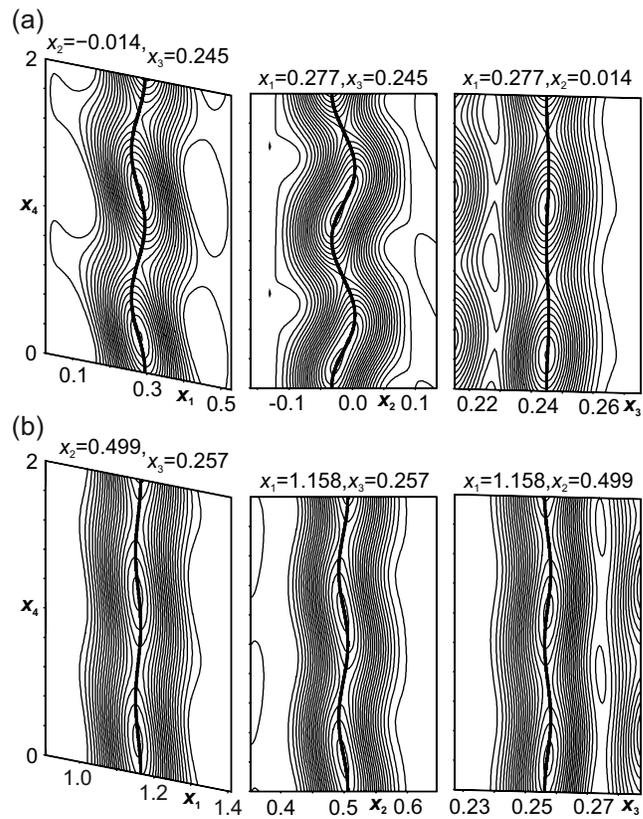}
\caption{($x_{si}$,$x_{s4}$)-sections of Fourier map centered on carbon atoms (black)
(a) C23a of molecule `A' and
(b) C23b of molecule `B' respectively.
The contour line and the width of the maps are 0.5 e{\AA}$^{-3}$ and 2.5 {\AA} respectively.
\label{l:biphome_C23_fourier}}
\end{figure}

The crystal structure of the room temperature phase (phase I hereon) was
redetermined at $T$ = 160 K using \textsc{Superflip}
\cite{palatinus_l_2007}
and refined using \textsc{Jana2006}
\cite{petricek_v_2014}
and \textsc{Jana2020}
\cite{petricek_v_2022}.
Atoms were renamed with suffixes `a' and `b'
for the two independent molecules `A' and `B'
[Fig.~\ref{l:biphome_160_labels_packing}(a)].
Anisotropic atomic displacement parameters (ADPs) of all non-hydrogens atoms were refined.
Hydrogen atoms (H) were added to carbon and nitrogen atoms
using riding model in ideal chemical geometry with
constraints for isotropic ADPs
[$U_{iso}$(H) = 1.2$U_{eq}$(N),
$U_{iso}$(H) = 1.2$U_{eq}$(C$_{\mathrm{aromatic}}$)
and $U_{iso}$(H) = 1.5$U_{eq}$(C$_{sp3}$)].
Since ${\beta}$ is close to 90 deg,
the integrated data was tested for twinning
employing twofold rotation parallel $\mathbf{a}$
as twin law.
This twin law is a true symmetry element in case
of a hypothetical orthorhombic lattice
with point group symmetry $222$
\cite{petricek_v_2016,Nespolo_M_2019}.
The fit of the structure model improved
(compare $R_{F}^{obs}$ = 0.0463 to 0.0408)
and volume of the second component refined to $\sim$ 2\%
(Table S2 in supporting information).
Finally, positions of the H atoms of NH groups and parameter corresponding to
isotropic extinction correction was refined
that further improved $R_{F}^{obs}$ values
(= 0.0393, Table S2 in supporting information).
The crystal structure reproduced the values for
intramolecular rotations reported those for
the structure at $T$ = 200 K
[$\varphi_{\mathrm{chiral}}$ = $\varphi_{1}$ (hereon) and $\psi$ in Fig.~\ref{l:biphome_160_labels_packing}(a)].
In addition, we also observe that the coplanar biphenyl rings are significantly
rotated with respect to the amide groups
[at $T$ = 200 K: $\varphi_{2}$ = 32.8 deg and 31.2 deg \cite{sasmal_2019_2},
at $T$ = 160 K in Fig.~\ref{l:biphome_160_labels_packing}(a)]
which also remains invariant as function of temperature.
The modulated structure of phase II at $T$ = 100 K
was refined using \textsc{Jana2006} and \textsc{Jana2020}.
Fractional co-ordinates of all non-hydrogen atoms
from the crystal structure at $T$ = 160 K
were used as starting model and the
average structure was refined against main reflections.
In successive steps, an incommensurate (IC) model described by
one harmonic wave for displacive modulation describing the atomic modulation functions (AMFs)
and basic parameters for anisotropic ADPs for all atoms
was refined against main and satellite reflections that
resulted in good fit to the diffraction pattern ($R_{F}^{obs}$ = 0.0425).
However, ADPs of four non-hydrogen atoms were found to be non-positive definite.
Since the components of $\mathbf{q}$, ${\sigma}_{1}$ and ${\sigma}_{3}$ are rational,
three commensurately modulated structures were pursued
by fixing the initial phase of the modulation to values,
$t_{0}$ = $0$,  $\frac{1}{4}$ and $\frac{1}{8}$ respectively.
While the former two $t_{0}$ values describe monoclinic $B2_{1}$ space group symmetry
for the equivalent 3D $2a\times{b}\times2c$ superstructure,
the third corresponds to triclinic $B1$ symmetry.
The commensurately modulated structure (C) model corresponding to $t_{0}$ = $\frac{1}{4}$
resulted in the best fit to the diffraction data ($R_{F}^{obs}$ = 0.0426)
including ADPs of all atoms positive definite.
As the atomic modulation functions (AMFs) have sinusoidal character,
the residual values are similar to the IC model
(Fig.~\ref{l:biphome_C23_fourier}, Fig. S2-S4 and
Table S2 in supporting information)
However, the C model at $t_{0}$ = $\frac{1}{4}$ is described
with either cosine or sine waves for the AMFs
(equal to number of refinable fractional coordinates in the equivalent superstructure)
reducing significantly the number of refinable parameters as compared to the IC model
(compare $N_{\mathrm{C}}$ = 649 to $N_{\mathrm{IC}}$ = 811, further tests in supporting information).
The final C model was further improved by refining the parameter corresponding to isotropic extinction correction and
AMFs and positions of hydrogen atoms of NH groups
($R_{F}^{obs}$ = 0.0419, Table S2 in supporting information).

\section{Results and discussion}

\subsection{Structural phase transition and unequal distortion of molecules}

\begin{figure}
\includegraphics{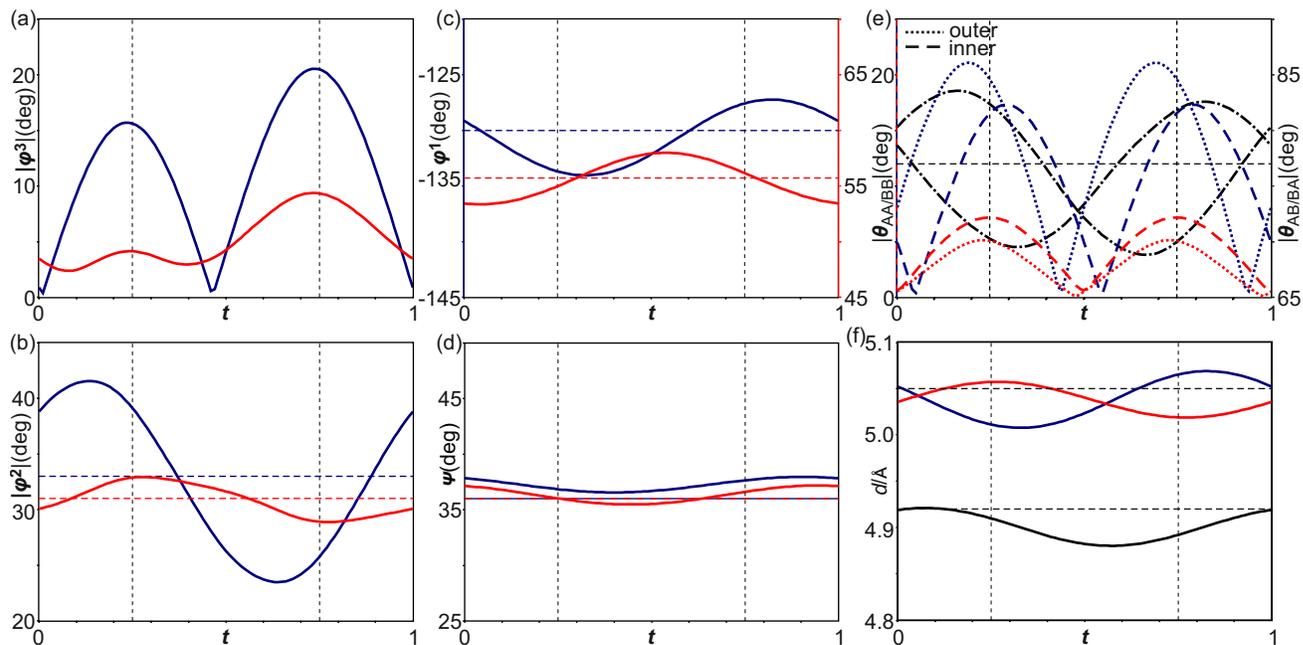}
\caption{$t$-plots of intramolecular rotations of molecule `A' (blue) and `B' (red)
as well as intermolecular tilts and distances between stacks
of 4-biphenylcarxy-(L)-phenylalaninate.
(a) Dihedral angle, ${\mid}{\varphi}^{3}{\mid}$ represents internal torsion
within biphenyl;
(b) Dihedral angle, ${\mid}{\varphi}^{2}{\mid}$ represents torsion
between the inner ring of biphenyl and the amide groups;
(c) ${\varphi}^{1}$ represents torsion of the amide groups with respect to the -COOCH$_{3}$
groups;
(d) ${\psi}$ represent torsion of -COOCH$_{3}$ groups with respect to amide groups.
(e) ${\mid}\theta_{\mathrm{AA/BB}}{\mid}$ represent tilt between biphenyl rings of `A' and `A'$^{ii}$ (blue); and of `B' and `B'$^{ii}$ (red)
and ${\mid}\theta_{\mathrm{AB/BA}}{\mid}$ (dashed-dotted black curve) represent tilt between inner aromatic rings of biphenyl (bonded to amide groups)
of `A' and outer ring of `B' and vice-versa.
(f) Intermolecular distances ($d$) between biphenyl rings of `A' and `A'$^{ii}$ (blue), between those of `B' and `B'$^{ii}$ (red)
and between those of `A' and `B' (black).
Horizontal dashed lines represent those angles and distances in phase I (${\mid}{\varphi}^{3}{\mid}$ = ${\mid}{\theta}_{AA/BB}{\mid}$ = 0 deg).
Vertical dashed lines indicate $t$ values corresponding to angles and distances in the 3D superstructure.
\label{l:biphome_intra_inter_torsion_dimer_t_plot}}
\end{figure}

\begin{figure}
\includegraphics{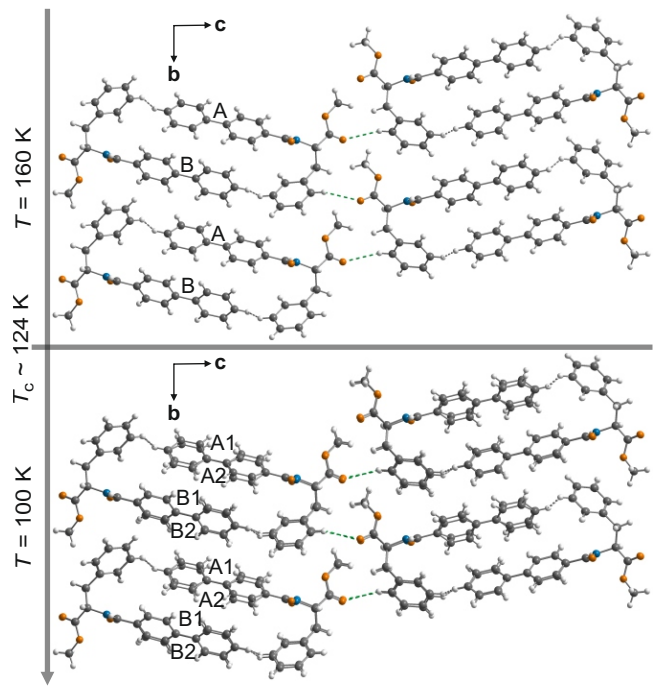}
\caption{Comparison of structures in phase I and phase II across the phase transition
highlighting the effect of internal torsion ($\varphi^{3}$) within biphenyl on the stacking arrangements along $\mathbf{a}$.
The tilt between the biphenyl stacks, $\theta_{\mathrm{AA/BB}}$
are different for the inner rings (bonded to amide rings) and the outer rings
[corresponding values in $t$-plot in Fig.~\ref{l:biphome_intra_inter_torsion_dimer_t_plot}(e)].
See full unit cells in Fig. S6 in supporting information.
View along [$\overline{1}$00].
\label{l:biphome_interaction_phase_transition}}
\end{figure}

In the present case, the monoclinic symmetry is retained below $T_{c}$
unlike monoclinic to triclinic distortion at disorder--order
phase transition of $p$-terphenyl
\cite{rice_ap_2013}
and $p$-quarterphenyl
\cite{baudour_jl_1978}.
In the final commensurately modulated structure model with $t_{0}$ = $\frac{1}{4}$,
sections corresponding to $t$ = $\frac{1}{4}$ and $\frac{3}{4}$
(Fig.~\ref{l:biphome_intra_inter_torsion_dimer_t_plot}, Fig. S5 in supporting information)
are physically relevant that corresponds to atomic positions in the equivalent
twofold superstructure in 3D
(Fig.~\ref{l:biphome_interaction_phase_transition}, Fig. S6 in supporting information).
Crystal structures of phase I and phase II
have group-subgroup relations
and the doubling of the $a$ and $c$ axes describes the additional
$B$-centering of the superstructure in II.
The superstructure derived using \textsc{Jana2006}
comprises of four molecules in the asymmetric unit ($Z^{\prime}$ = 4);
two each corresponding to molecule `A' and `B' of phase I
(Fig.~\ref{l:biphome_interaction_phase_transition}).
The covalent bond distances are similar for the independent
set of molecules and are practically unaffected by modulation
(Table S6 in supporting information).
In the present study, discussion is based on the modulated structure
in order to establish unique relations between phase I and II respectively
\cite{rekis_t_2021,ramakrishnan_s_2019,dey_s_2016,noohinejad_l_2015,schoenleber_a_2011,schoenleber_a_2003}.
The modulated structure suggests that the phase transition is
dominated by evolution of internal torsional degrees of freedom
(${\varphi}^{3}$ $>$ $0^{\circ}$)
within the biphenyl moieties
[Fig.~\ref{l:biphome_intra_inter_torsion_dimer_t_plot}(a)].
The twists about the central C--C bond are significantly different
for the two molecules where the torsional modulation of
`A' are 2-4 times larger than those of `B'
(dihedral angle: $\mid{\varphi}_{A}^{3}\mid$ = 15.6 deg, 20.5 deg; $\mid{\varphi}_{B}^{3}\mid$ = 4.1 deg, 9.3 deg).
These torsions are described by highly anisotropic AMFs ($u$) along the three basis vectors
where the maximum amplitude are along $\mathbf{b}$ for the carbon atoms of biphenyl
(Fig.~\ref{l:biphome_C23_fourier} and Table S4 in supporting information).
Notably, the rotations in the present structure are significantly larger
than those reported for molecular biphenyl
[$\varphi$ $\simeq$ $\pm$ 5.5 deg in
ref.\cite{petricek_v_1985,baudour_jl_1983,baudour_jl_1983}]
but smaller than those in the low temperature superstructure
of $p$-terphenyl and $p$-quarterphenyl
[Maximum $\varphi_{\mathrm{terphenyl,quarterphenyl}}$ $\simeq$ 23 deg
in ref.\cite{rice_ap_2013,baudour_jl_1976,baudour_jl_1978}].
Another distinctive property of the modulated structure is the
unequal modulation for the two different moieties where
$u_{\mathrm{biphenyl}}$ $>$ $u_{\mathrm{phenylalaninate}}$
(Table S4 in supporting information).
Therefore, the variation in torsions $\varphi^{1}$ and $\psi$ are smaller
[Fig.~\ref{l:biphome_intra_inter_torsion_dimer_t_plot}(c),(d)].
A possible reason for the weaker modulations of the atoms around the chiral centers
is the directional strong intermolecular N--H${\cdot}{\cdot}{\cdot}$O
bonds makes large intramolecular rotations unfavorable.
Note that the observed changes in the rotations $\varphi^{2}$ [Fig.~\ref{l:biphome_intra_inter_torsion_dimer_t_plot}(b)]
of molecule `A'
are predominantly described by strong modulations of the molecule's biphenyl moiety.
The asymmetry in rotations of individual molecules 
($\Delta\mid{\varphi}^{3}\mid$ $\approx$ 5 deg)
is determined by the disparate bonding environments 
of the biphenyl moieties where the inner rings are that are covalently bonded to amide groups
while the outer interact weakly \textit{via} C--H${\cdot}{\cdot}{\cdot}$H--C
interactions with the phenyl rings of phenylalaninate groups (Fig.~\ref{l:biphome_interaction_phase_transition}).
Subsequently, the unequal values at the relevant $t$-sections of $\varphi^{3}$
are correlated to those of $\varphi^{2}$
[compare Fig.~\ref{l:biphome_intra_inter_torsion_dimer_t_plot}(a) and Fig.~\ref{l:biphome_intra_inter_torsion_dimer_t_plot}(b)]
suggesting the chemical influence of the amide groups on the phenyl rings and \textit{vice-versa}.

In the modulated structure,
the biphenyl moieties in (AA)$_{\mathrm{n}}$ and (BB)$_{\mathrm{n}}$
stacks are
tilted with respect to each other
[Fig.~\ref{l:biphome_intra_inter_torsion_dimer_t_plot}(e)]
which are parallel in phase I.
These tilts ($\theta_{\mathrm{AA/BB}}$)
are of the order of the internal twists,
$\varphi^{3}$
of the independent biphenyl moieties
[compare Fig.~\ref{l:biphome_intra_inter_torsion_dimer_t_plot}(e) to Fig.~\ref{l:biphome_intra_inter_torsion_dimer_t_plot}(a)]
The orientation between the biphenyl moieties
within the ({ABAB})$_{\mathrm{n}}$ stacks
also vary with $\Delta\theta_{\mathrm{AA/BB}}$ $\simeq$ 12 deg
where the value is intermediate to ${\varphi}_{A}^{3}$ and ${\varphi}_{B}^{3}$
[compare Fig.~\ref{l:biphome_intra_inter_torsion_dimer_t_plot}(e) to Fig.~\ref{l:biphome_intra_inter_torsion_dimer_t_plot}(a)].
In addition,
intermolecular distances between the biphenyl moieties within the stacks at the
two $t$-sections are different and vary up to $\Delta{d}_{\mathrm{AA/BB}}$ $\simeq$ 0.05 \AA
and $\Delta{d}_{\mathrm{ABAB}}$ $\simeq$ 0.02 \AA
[Fig.~\ref{l:biphome_intra_inter_torsion_dimer_t_plot}(f)].
However, these variations in $d$
are small compared to the molecular tilts, $\theta$.
It could therefore be argued that the dimerization of biphenyl molecular stacks below $T_{c}$ are
predominantly governed by distortion described by molecular rotations rather than
intermolecular distances.
On the other hand,
intermolecular distances between aromatic rings of L-phenylalaninate
vary similarly to the biphenyls
albeit significantly smaller interstack rotations, $\theta$ $<$ 3 deg
(Fig. S5 in supporting information).

\begin{figure}
\includegraphics{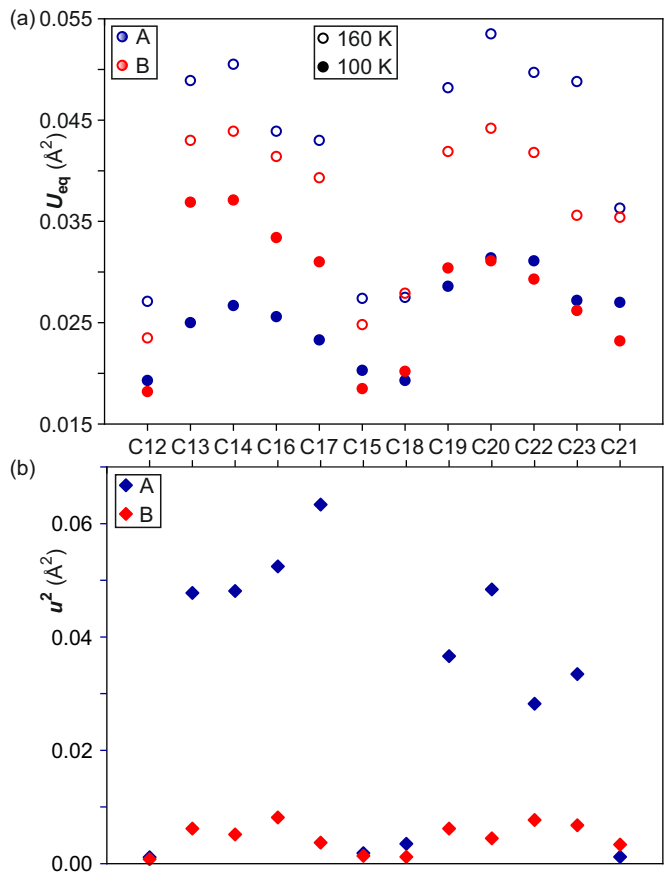}
\caption{Scatter plots of equivalent value of anisotropic ADPs ($U_{eq}$)
and square of the amplitude of modulations ($u^{2}$) of the carbon atoms of biphenyl moieties (C12 through to C23,
see Fig.~\ref{l:biphome_160_labels_packing}.)
of molecule `A' (blue) and `B' (red).
(a) $U_{eq}$ of the carbon atoms at $T$ = 160 K
(open circles) and at $T$ = 100 K (full circles).
(b) $u^{2}$ (diamonds) of the corresponding carbon atoms of molecule `A' and `B'.
See Table S5 in supporting information.
\label{l:biphome_adp_amf}}
\end{figure}

The increased distortions are also accompanied by
suppression of dynamic disorder below $T_{c}$.
For example, the carbon atoms at ortho (C14, C16, C19, C23)
and meta (C13, C17, C20, C22) positions are strongly displaced
(Fig.~\ref{l:biphome_adp_amf}, Table S4 and S5 in supporting information).
Subsequently, the ADPs are significantly reduced
as compared to phase I
(Fig.~\ref{l:biphome_adp_amf}, Table S5 in supporting information).
Notably, decrease of the ADPs ($U_{eq}$) from $T$ = 160 K to $T$ = 100 K
is larger for those of molecule `A' than those for `B',
while the square of the amplitude of modulations ($u^{2}$)
are greater for `A' than those for `B'
[compare Fig.~\ref{l:biphome_adp_amf}(b) to Fig.~\ref{l:biphome_adp_amf}(a)].

\subsection{Competitive forces governing modulations}

Structural studies in the 3D phase of molecular biphenyl have suggested
that the ortho-hydrogen atoms are displaced away
in the plane of the rings to minimize steric hindrance
\cite{trotter_j_1961,hargreaves_a_1962,charbonneau_gp_1976}.
On the other hand, dynamic disorder predominantly
governed by torsional vibrations around the long molecular axis
\cite{petricek_v_1985}
is predicted to balance the planar conformation of biphenyl
favourable for crystal packing
\cite{lenstra_ath_1994}.
As short as 1.98 {\AA} in phase I
(Table~\ref{tab:h_distances}), these contacts are
shorter than the predicted values for twice van der Waals
radius for hydrogen ($r$ = 1.1--1.2 {\AA} \cite{rowland_rs_1996,alvarez_s_2013}).
In the modulated structure, we observe that the distances between the ortho-hydrogen
atoms are marginally but consistently larger than those in phase I
(Table~\ref{tab:h_distances})
that could suggest that the torsional modulations aid in
minimization of the presumed steric hindrance below $T_{c}$
\cite{dey_s_2022,dey_s_2018}.

A peculiar property of the modulated structure under discussion
is the significant difference in the torsional amplitude ${\varphi}^{3}$
of the independent molecules.
This aspect cannot be explained solely based on the
intramolecular steric factors.
Analysis of the crystal packing shows that each of these independent biphenyl moieties
maintains close intermolecular CH${\cdot}{\cdot}{\cdot}$HC
contacts with the phenyl rings of L-phenylalaninate
in AB and BA fashion
(Fig.~\ref{l:biphome_interaction_phase_transition}).
These distances are significantly longer
(intermolecular $d_{\mathrm{H{\cdot}{\cdot}{\cdot}H}}$ $\geq$ 2.4 \AA,
Table~\ref{tab:h_distances})
compared to the intramolecular H${\cdot}{\cdot}{\cdot}$H distances.
On the other hand, the aromatic rings of L-phenylalaninate
interact with adjacent oxygen atoms of
--COOCH$_{3}$ \textit{via} C--H${\cdot}{\cdot}{\cdot}$O hydrogen bonds
(Fig.~\ref{l:biphome_interaction_phase_transition} and
Table~\ref{tab:h_distances}).
These hydrogen bonds are weaker \cite{desiraju_gr_2001}
but highly directional
($\angle$(C--H${\cdot}{\cdot}{\cdot}$O) = 159--164 deg)
with very little variation in the distances.
Interestingly, those H${\cdot}{\cdot}{\cdot}$H distances involving
biphenyl moieties of `B' are consistently
smaller than those of `A' in both phases
(Table~\ref{tab:h_distances}).
We argue that in the presence of both the van der Waals interactions
and weak C--H${\cdot}{\cdot}{\cdot}$O bonds,
the larger distortions of `A' is favored
by weaker CH${\cdot}{\cdot}{\cdot}$HC interactions
while that is suppressed in `B'.
The four different values of intramolecular torsion $\varphi^{3}$
within the biphenyl moieties is distinctively
governed by intra- and intermolecular non-bonded H${\cdot}{\cdot}{\cdot}$H contacts
as well as weak hydrogen bonds.

\section{Conclusions\label{conc}}

The single crystal to single crystal phase transition
of 4-biphenylcarboxy-(L)-phenylalaninate below $T$ = 124 K
drives the 3D structure directly to a locked-in
twofold superstructure.
The commensurately modulated structure
at $T$ = 100 K is accompanied by
pronounced amplitudes of torsion within
biphenyl that are characteristic of modulated
and superstructures of other polyphenyls.
The phase transition temperature is significantly higher than
that in biphenyl yet significantly smaller than for p-terphenyl
and p-quarterphenyl.
Consistent with the $T_{c}$,
the maximum amplitude of torsion
is also intermediate and in the order
${\varphi}_{\mathrm{quarterphenyl}}$ $\geq$ ${\varphi}_{\mathrm{terphenyl}}$ $>$ ${\varphi}_{\mathrm{4-biphenylcarboxy-L-phenylalaninate}}$ $>$ ${\varphi}_{\mathrm{biphenyl}}$.

Topologically separated, conformations of both the weaker C--H${\cdot}{\cdot}{\cdot}$O bonds
and stronger N--H${\cdot}{\cdot}{\cdot}$O bonds are rigid and
that underlines their role in stabilizing the crystal packing in both phases.
A unique property of the present polyphenyl coupled amino acid ester
is the distinctively unequal torsional amplitude
(${\varphi}_{A}$ $>$ ${\varphi}_{B}$)
within the independent molecules
which is governed by multiple level of competitions involving unequal van der Waals constraints
in presence of weak hydrogen bonds between the biphenyl
and L-phenylalaninate moieties.
The unusual nature of the phase transition
is described by the fact that unequal displacive modulations of the
two molecules are complimented by unequal suppression
of the dynamic disorder of their atoms below $T_{c}$.
This difference in the torsion as well as the phase transition conditions
highlights
how conjugation of polyphenyls can possibly
be influenced by amino acid esters
which in turn influences
supramolecular assemblies.

     % Appendices appear after the main body of the text. They are prefixed by
     % a single \appendix declaration, and are then structured just like the
     % body text.

%\appendix
%\section{Appendix title}
%
%Text text text text text text text text text text text text text text
%text text text text text text text.
%
%\subsection{Title}
%
%Text text text text text text text text text text text text text text
%text text text text text text text.
%
%\subsubsection{Title}
%
%Text text text text text text text text text text text text text text
%text text text text text text text.
%
%
%     %-------------------------------------------------------------------------
%     % The back matter of the paper - acknowledgements and references
%     %-------------------------------------------------------------------------
%
%     % Acknowledgements come after the appendices

\ack{Acknowledgements}
We thank Prof. Sreenivasan Ramakrishnan, Dr. Vaclav Petricek, Dr. Sitaram Ramakrishnan,
Prof. Venkataramanan Mahalingam and Dr. Saumya Mukherjee
for helpful comments and fruitful discussions.
Financial support from SERB-DST (DST-SERB:PDF/2018/002502) and Alexander von Humboldt foundation
is gratefully acknowledged.

\clearpage

 \begin{table}
 \caption{Temperature dependence of lattice parameters and the components of the modulation wave vector,
 $\sigma_{1}$ and $\sigma_{3}$. See Table S1 in supporting information for reflections used.
 \label{tab:T_SCXRD}}
 \begin{tabular}{llllllll}
 $T$ (K) & $a$ (\AA) & $b$ (\AA) & $c$ (\AA) & $\beta$ (deg) & $\sigma_{1}$ & $\sigma_{3}$ & $V$ (\AA$^{3}$) \\
 \hline
 200\cite{sasmal_2019_2} & 5.0560(3) & 8.6622(4) & 42.242(3) & 90.349(4) & & & 1850.00(18) \\
 160 & 5.0479(2) & 8.6330(4) & 42.1525(15) & 90.513(3) & & & 1836.87(13)  \\
 150 & 5.0498(7) & 8.6161(8) & 42.136(11) & 90.607(11) & & & 1833.2(5) \\
 140 & 5.0484(6) & 8.6093(7) & 42.145(10) & 90.661(10) & & & 1831.6(5) \\
 130 & 5.0451(7) & 8.6014(8) & 42.120(11) & 90.713(11) & & & 1827.6(6) \\
 128 & 5.0446(7) & 8.6002(8) & 42.113(11) & 90.723(11) & & & 1826.9(6) \\
 126 & 5.0440(6) & 8.5992(7) & 42.114(10) & 90.720(9) & & & 1826.5(5) \\
 124 & 5.0443(6) & 8.5978(7) & 42.100(10) & 90.733(9) & 0.499(9)& 0.51(7) & 1825.7(5) \\
 122 & 5.0440(7) & 8.5970(7) & 42.100(10) & 90.745(10) & 0.498(6) & 0.49(4) & 1825.5(5) \\
 120 & 5.0433(7) & 8.5984(7) & 42.100(10) & 90.746(10) & 0.497(5) & 0.52(3) & 1825.5(5) \\
 118 & 5.0441(7) & 8.5958(8) & 42.092(11) & 90.764(10) & 0.500(4) & 0.50(3) & 1824.8(5) \\
 116 & 5.0432(6) & 8.5926(7) & 42.087(9) & 90.775(9) & 0.500(4) & 0.51(3) & 1823.6(5) \\
 114 & 5.0422(6) & 8.5923(7) & 42.090(10) & 90.787(10) & 0.500(4) & 0.53(3) & 1823.3(5) \\
 100 & 5.0377(2) &  8.5898(3)&  42.0432(14) & 90.884(3) &  0.5 & 0.5 & 1819.11(11) \\
 \hline
 \end{tabular}
 \end{table}
 
 \begin{table}
 \caption{Comparison of non-bonded hydrogen${\cdot}{\cdot}{\cdot}$acceptor and
 hydrogen${\cdot}{\cdot}{\cdot}$hydrogen distances (\AA) involved in
 hydrogen bonds and steric factors in phase I ($T_{1}$)
 and phase II ($T_{2}$; $t$ = $\frac{1}{4}$; $\frac{3}{4}$) respectively.
 Symmetry codes: Phase I-- (i) $x-1,y,z$; (ii) $x+1,y,z$; (iii) $-x+1,y+\frac{1}{2},-z+1$; (iv) $-x+2,y-\frac{1}{2},-z$.
 Phase II-- (i) $x-1,y,z,t$; (ii) $x+1,y,z,t$; (iii) $-x+1,y+\frac{1}{2},-z+1,-t$; (iv) $-x+2,y-\frac{1}{2},-z,-t$.
 \label{tab:h_distances}}
 \begin{tabular}{llll}
 Interaction class               & Atom group labels & Phase & distances (\AA) \\
 \hline
 N--H${\cdot}{\cdot}{\cdot}$O    & H1n1a${\cdot}{\cdot}{\cdot}$O3a$^{i}$       & I  & 2.00       \\
                                 &                                             & II & 1.97; 1.95 \\
                                 & H1n1b${\cdot}{\cdot}{\cdot}$O3b$^{ii}$      & I  & 2.00       \\
                                 &                                             & II & 2.00, 1.98 \\
 C--H${\cdot}{\cdot}{\cdot}$O    & H1c10a${\cdot}{\cdot}{\cdot}$O2a$^{iii}$    & I  & 2.72       \\
                                 &                                             & II & 2.72; 2.68 \\
                                 & H1c10b${\cdot}{\cdot}{\cdot}$O2b$^{iv}$     & I  & 2.78       \\
                                 &                                             & II & 2.74; 2.73 \\
 Intra H${\cdot}{\cdot}{\cdot}$H & H1c14a${\cdot}{\cdot}{\cdot}$H1c19a         & I  & 1.98       \\
                                 &                                             & II & 2.08; 2.09 \\
                                 & H1c14b${\cdot}{\cdot}{\cdot}$H1c19b         & I  & 1.98       \\
                                 &                                             & II & 2.00; 1.99 \\
                                 & H1c16a${\cdot}{\cdot}{\cdot}$H1c23a         & I  & 2.03       \\
                                 &                                             & II & 2.10; 2.16 \\
                                 & H1c16b${\cdot}{\cdot}{\cdot}$H1c23b         & I  & 2.02       \\
                                 &                                             & II & 2.03; 2.06 \\
 Inter H${\cdot}{\cdot}{\cdot}$H & H1c21a${\cdot}{\cdot}{\cdot}$H1c7b          & I  & 2.59       \\
                                 &                                             & II & 2.61; 2.56 \\
                                 & H1c21b${\cdot}{\cdot}{\cdot}$H1c7a$^{ii}$   & I  & 2.45       \\
                                 &                                             & II & 2.39; 2.46 \\
 \hline
 \end{tabular}
 \end{table}
 

     % References are at the end of the document, between \begin{references}
     % and \end{references} tags. Each reference is in a \reference entry.

%\begin{references}
%\reference{Author, A. \& Author, B. (1984). \emph{Journal} \textbf{Vol},
%first page--last page.}
%\end{references}

     %-------------------------------------------------------------------------
     % TABLES AND FIGURES SHOULD BE INSERTED AFTER THE MAIN BODY OF THE TEXT
     %-------------------------------------------------------------------------

     % Simple tables should use the tabular environment according to this
     % model

%\begin{table}
%\caption{Caption to table}
%\begin{tabular}{llcr}      % Alignment for each cell: l=left, c=center, r=right
% HEADING    & FOR        & EACH       & COLUMN     \\
%\hline
% entry      & entry      & entry      & entry      \\
% entry      & entry      & entry      & entry      \\
% entry      & entry      & entry      & entry      \\
%\end{tabular}
%\end{table}

     % Postscript figures can be included with multiple figure blocks

%\begin{figure}
%\caption{Caption describing figure.}
%\includegraphics{fig1.ps}
%\end{figure}

\referencelist[biphome_dey]

\end{document}

%%%%%%%%%%%%%%%%%%%%%%%%%%%%%%%%%%%%%%%%%%%%%%%%%%%%%%%%%%%%%%%%%%%%%%%%%%%%%%%
%%%%%%%%%%%%%%%%%%%%%%%%%%%%%%%%%%%%%%%%%%%%%%%%%%%%%%%%%%%%%%%%%%%%%%%%%%%%%%%

\begin{center}

  ~\\
  \bigskip

  {\LARGE{\textbf{Supporting information}}}\\
  \bigskip
  \bigskip
  \bigskip
  {\LARGE{\textbf{Asymmetric rotations and dimerization driven by normal to modulated phase transition in 4-biphenylcarboxy coupled L-phenylalaninate}}}

  \bigskip
  \bigskip
  \bigskip

  {{Somnath Dey$^{a,b*}$, Supriya Sasmal$^{a}$,
  Saikat Mondal$^{a}$, Santosh Kumar$^{a}$, Rituparno Chowdhury$^{a}$, Debashrita Sarkar$^{a}$,
  C. Malla Reddy$^{a}$, Lars Peters$^{b}$, Georg Roth$^{b}$ and Debasish Haldar$^{a}$}}

  \bigskip
  \bigskip

  {\textit{$^{a}$Department of Chemical Sciences, Indian Institute of Science Education and Research (IISER) Kolkata,
  Mohanpur 741246, India,\\
  $^{b}$Institute of Crystallography, RWTH Aachen University, Jägerstraße 17-19,
  52066 Aachen, Germany}\\
    ~\\
  E-mail: dey@ifk.rwth-aachen.de}

\end{center}

\bigskip

%--------------------

\section*{Contents}

\smallskip
\noindent

\noindent Details of structure refinements
\smallskip
\newline \noindent Powder X-ray diffraction experiments
\smallskip
\newline \noindent Supplementary Figures S1--S7
\smallskip
\newline \noindent Supplementary Table S1--S7.

\clearpage

\subsection*{Details of structure refinements of different modulated structure models}

Structure refinements have been performed using
\textsc{Jana2006}
\cite{petricek_v_2014}
and \textsc{Jana2020}
\cite{petricek_v_2022}.
Structural model at $T$ = 160 K has been used
as an initial model for the basic structure of the modulated structure
at $T$ = 100 K.
All atoms were set to isotropic for displacement parameters
and the model was refined against main reflections
[$R_{F}^{obs}$($m$=0) = 0.0723].
In the next step, first order harmonic for
displacive modulation was described for all atoms
and an incommensurate (IC) model was refined against
main and satellite reflections.
Refinement led to improved fit
to the main reflections
[$R_{F}^{obs}$($m$=0) = 0.0586, $R_{F}^{obs}$($m$=1) = 0.1215].
Refinement of the anisotropic atomic displacement parameters (ADPs)
of all non-hydrogen atoms resulted in
significant improvement to the residual values
[IC model A: $R_{F}^{obs}$($m$=0) = 0.0374, $R_{F}^{obs}$($m$=1) = 0.0771]
and residual features
(${\Delta}{\rho}_{min}$/${\Delta}{\rho}_{max}$)
decreased from
-0.68/1.23 e{\AA}$^{-3}$ to -0.33/0.31 e{\AA}$^{-3}$.
However, ADPs of four non-hydrogen atoms were
found to be non-positive definite.
Further test by describing
first order harmonic for
ADP modulation for all non-hydrogen atoms model
led to improvement of the residual values
[IC model A: $R_{F}^{obs}$($m$=0) = 0.0363, $R_{F}^{obs}$($m$=1) = 0.0677]
but ADPs of 11 non-hydrogen atoms were found to be non-positive definite
along certain $t$-sections.
This model was discarded for further analysis.

In the next step, IC model A was used as a starting model
to describe three
commensurate (C) models by fixing the initial phase of the modulation,
$t_{0}$ = $0$,  $\frac{1}{4}$ and $\frac{1}{8}$ respectively.
The former two $t_{0}$ values correspond to
$2a\times{b}\times2c$
superstructure in 3D with monoclinic symmetry $B2_{1}$ while the later
correspond to a superstructure with triclinic $B1$ symmetry.
Restrictions on $t_{0}$ values also impose
constraints on the refinable variables corresponding to atomic modulation functions (AMFs).
These restrictions follow the argument that the total number of refinable parameters in the
equivalent 3D superstructure and their (3+1)D commensurately modulated structural models
must be equal.
In the present case, either $\sin$ or $\cos$ waves can be refined for structural models with $t_{0}$ = 0 and $\frac{1}{4}$
because the point group symmetry is same in their corresponding 3D superstructure models.
On the other hand, assumed monoclinic to triclinic distortion
in the 3D superstructure (space group $B1$)corresponding to (3+1)D C model with $t_{0}$ = $\frac{1}{8}$
can be derived by using both components of the
Fourier series.
It must be noted that such restrictions on $\sin$ and $\cos$ waves
cannot be formally imposed on the AMFs of hydrogen atoms in
\textsc{Jana2006} and \textsc{Jana2020}
as their modulations are fully determined by geometrical conditions
of the riding model.
$t_{0}$ = $\frac{1}{4}$ yielded the best fit to the diffraction data
(Table~\ref{t:biphome_partial_r_values_t0_sup})
with reduced number of parameters as compared to the IC model A
(compare $N_{C,t_{0}=0.25}$ = 649 to $N_{IC,model\,A}$ = 811).
Most importantly, ADPs of all non-hydrogen atoms are positive definite.

Notably, the residual values of the IC model as well as
the C model at $t_{0}$ = $\frac{1}{8}$ is marginally smaller
than for the C model at $t_{0}$ = $\frac{1}{4}$.
Assuming all the three models should fit similarly to the diffraction data
for equivalent descriptions of structural models
further tests included attempts to refine the IC model
and C model at $t_{0}$ = $\frac{1}{8}$
with reduced number of parameters (= 649) similar to $t_{0}$ = $\frac{1}{4}$.
Refinements led to worse fit with large $R$-values
(Table~\ref{t:biphome_partial_r_values_t0_sup}).

In the final step,
all reflections were averaged in monoclinic symmetry
corresponding to $t_{0}$ = 0.
One parameter corresponding to isotropic extinction correction was refined.
Finally, fractional co-ordinates and AMFs for hydrogen atoms belonging to
N--H groups involved in strong hydrogen bonds
improved the fit to the diffraction data marginally
($R_{F}^{obs}$ = 0.0419 in Table~\ref{t:experimental and crystallographic table}).

Additional refinement was performed including first order harmonic
for anisotropic ADPs of all non-hydrogen atoms.
Refinement of this model with additional 324 parameters
converged with marginal improvement
of $R_{F}^{obs}$ (= 0.0406) values.
However, the residual density
${\Delta}{\rho}_{min}$/${\Delta}{\rho}_{max}$ remained
unchanged [compare -0.26/0.28 e/\AA{}$^{3}$ to -0.25/0.29 e/\AA{}$^{3}$]
and 306 parameters refined to values within three times their
standard uncertainties.
The model was therefore discarded.
Thus the superspace approach reduced the
total numbers of refinable parameters by $\sim$ 33 \%.

\subsection*{Additional X-ray diffraction experiments}

\begin{figure}
\begin{center}
\includegraphics{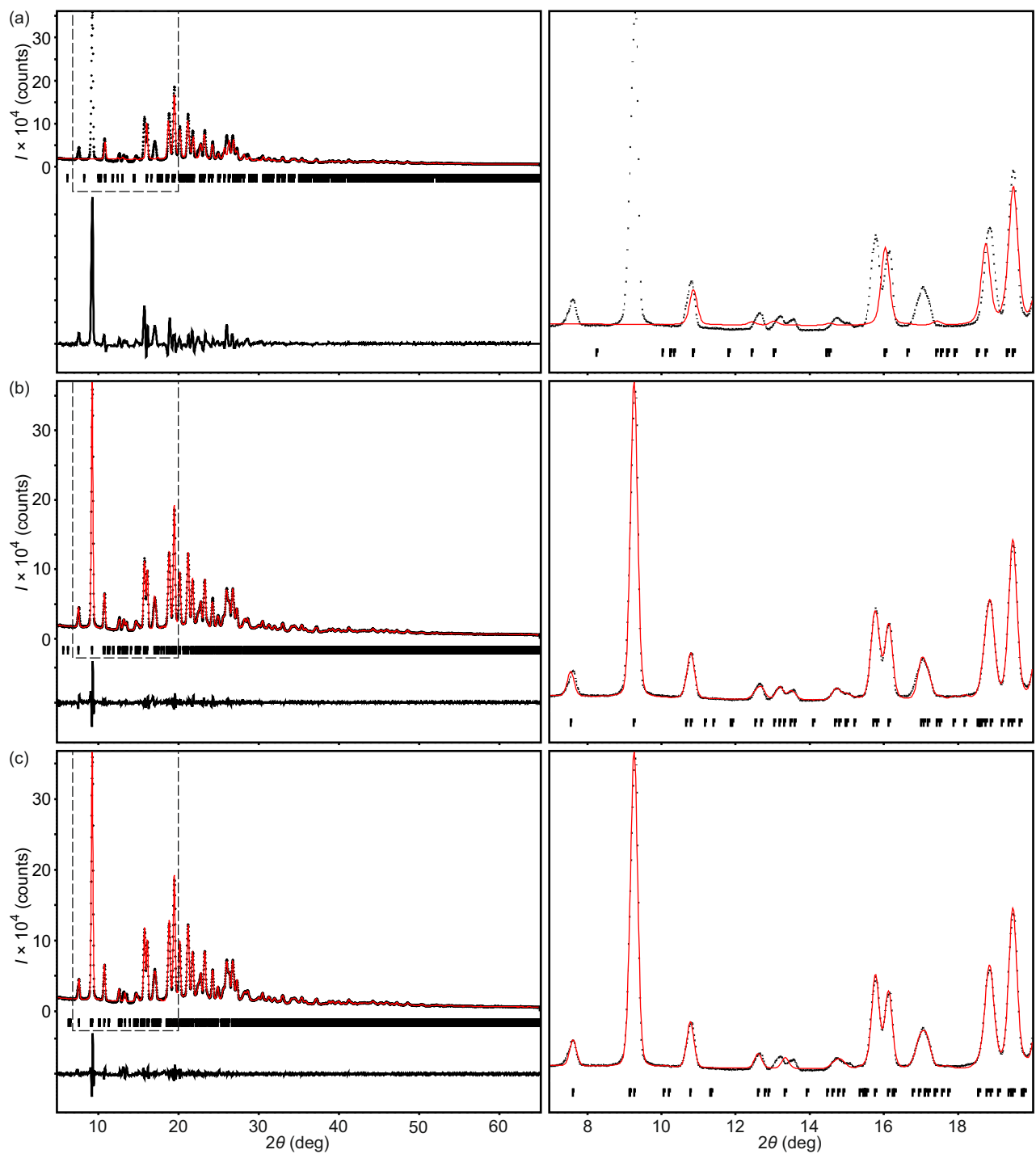}
\caption{Comparison of fit of the experimental powder X-ray diffraction pattern to
(a) as obtained unit cell from SCXRD, (b) Calculated unit cell 1 with volume 1901 {\AA}$^{3}$ and
(c) calculated unit cell with volume 2029 {\AA}$^{3}$.
Experimental pattern, calculated profile and difference are given in black cross points, red curve and black curve respectively.
The insets in 2$\theta$ = 7-20 deg are given in the right column corresponding to the area (dashed rectrangle) in left.}
\label{f:sup_biphome_pxrd}
\end{center}
\end{figure}

Powder X-ray diffraction experiments were performed
on thoroughly ground powder of the compound
at ambient conditions using a Rigaku SmartLab with a CuK$\alpha$ radiation.
\textsc{JANA2006} was used to index the diffraction patterns.
For reference, lattice parameters at ambient conditions were
obtained from single crystal X-ray diffraction (SCXRD)
experiment at ambient conditions (Table~\ref{t:lat_par_compare}).
The PXRD pattern
could not be indexed using the lattice parameters
as obtained from the SCXRD data
[Fig.~\ref{f:sup_biphome_pxrd}(a)] that suggest that the compound undergoes phase
transition upon grinding.
Lattice parameters were calculated employing
the singular value decomposition (SVD)-Index algorithm
in \textsc{TOPAS}
\cite{coelho_aa_2018,coelho_aa_2003}.
The PXRD pattern could be indexed using a primitve triclinic cell (Cell 1)
with unit cell volume comparable to that of single crystal
[Table~\ref{t:lat_par_compare}, Fig ~\ref{f:sup_biphome_pxrd}(b)].
Another triclinic cell (Cell 2) could also describe the pattern [Fig.~\ref{f:sup_biphome_pxrd}(c)].
Le Baile refinements of the patterns against both the cells resulted in similar
residual values (Table~\ref{t:lat_par_compare}).
However, Cell 1 fits better to the PXRD than Cell 2
[compare inset plots of Fig ~\ref{f:sup_biphome_pxrd}(b) and Fig ~\ref{f:sup_biphome_pxrd}(c)].
In addition, the unit cell volume of Cell 2 is larger than 7.5 \%
to that of the single crystal
that implies different density of the ground material.

Based on this difference of phases between single crystals (monoclinic structure)
and pulverised material (triclinic structure),
$T$-dependent PXRD
experiments to complement the
single crystal to single crystal phase transition
in this material was not pursued.

\clearpage

\begin{figure}
\begin{center}
\includegraphics{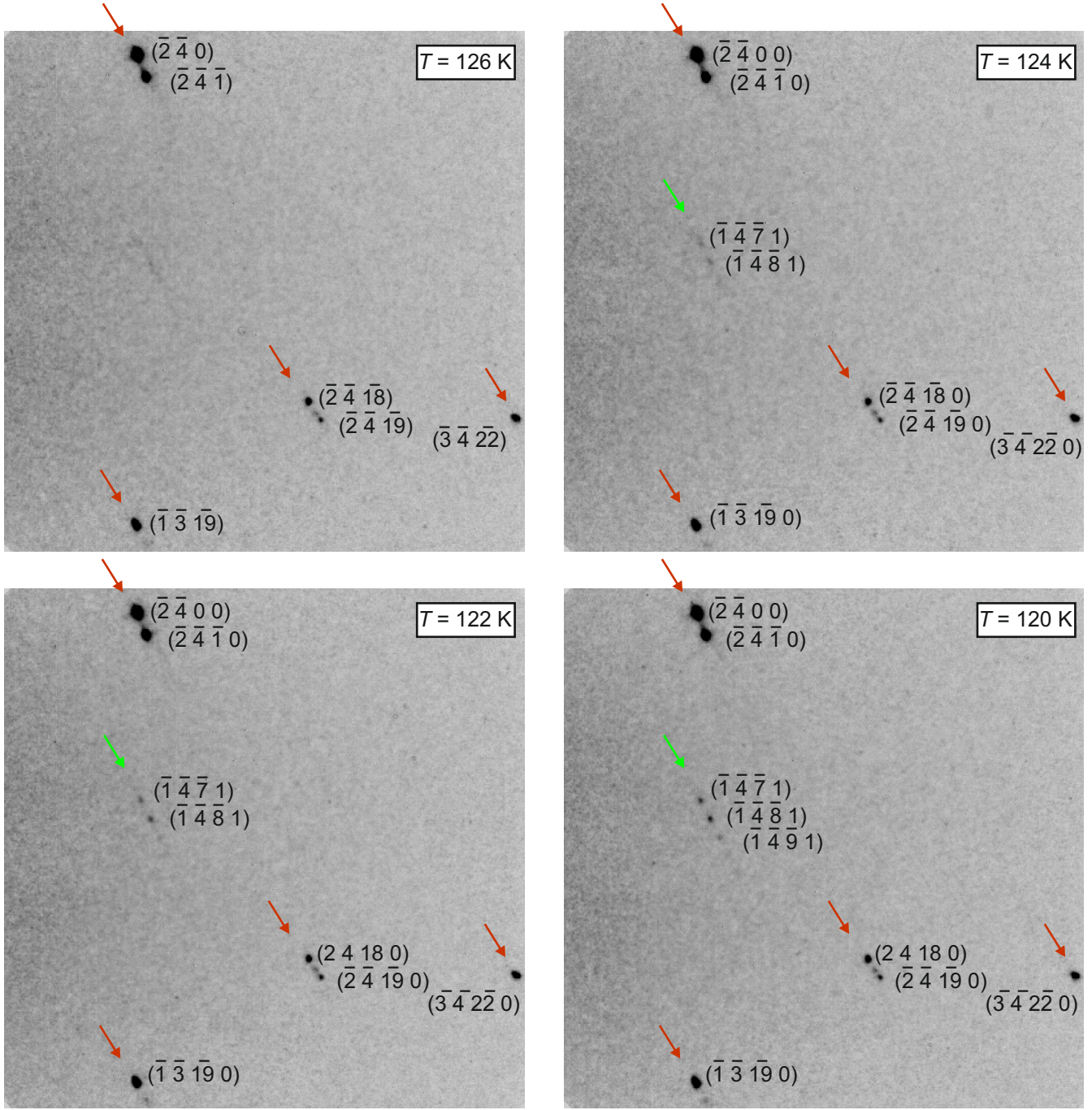}
\caption{Diffraction images across the normal (phase I) to commensurately modulated (phase II) phase transition.
Red arrows depict Bragg peaks in phase I and main Bragg peaks in phase II.
The satellites are diffuse at $T$ = 124 K (green arrow) that becomes stronger
at lower temperatures.
Reflections at $T$ = 126 K are indexed using three integers ($hkl$)
and at $T_c$ = 124 K and lower temperatures by four integers ($hklm$), where
$m$ = 0 and $m$ = 1 for main and satellite reflections respectively.
Image resolution range in $d$ $\sim$ 2.2 to 1.1 {\AA}.}
\label{f:sup_biphome_T_diffraction}
\end{center}
\end{figure}
\begin{figure}
\begin{center}
\includegraphics{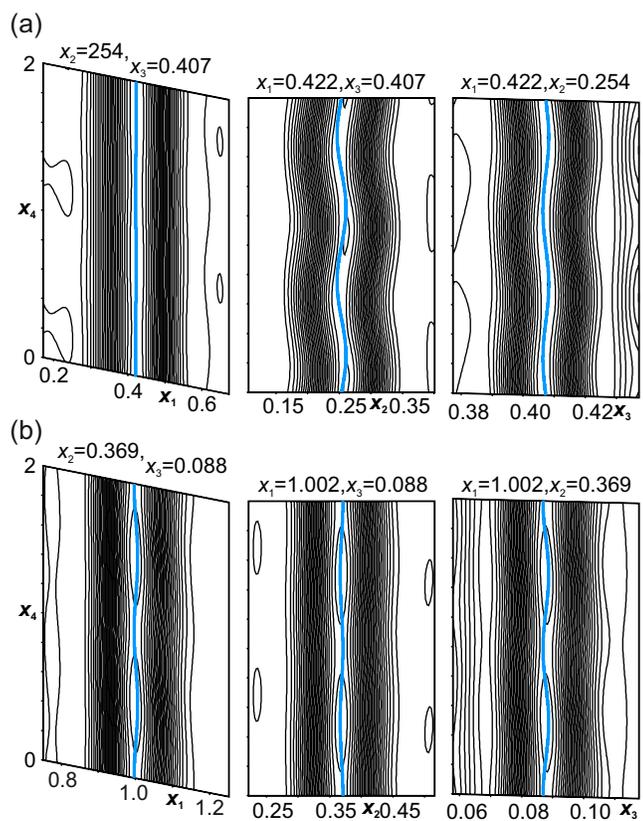}
\caption{($x_{si}$,$x_{s4}$)-sections of Fourier map centered on nitrogen atoms (light blue) of amide groups
(a) atom N1a of molecule `A' and (b) atom N1b of molecule `B'.
The contour line and the width of the maps are 0.5 e{\AA}$^{-3}$ and 2.5 {\AA} respectively.
\label{l:sup_biphome_N1_fourier}}
\end{center}
\end{figure}
\begin{figure}
\begin{center}
\includegraphics{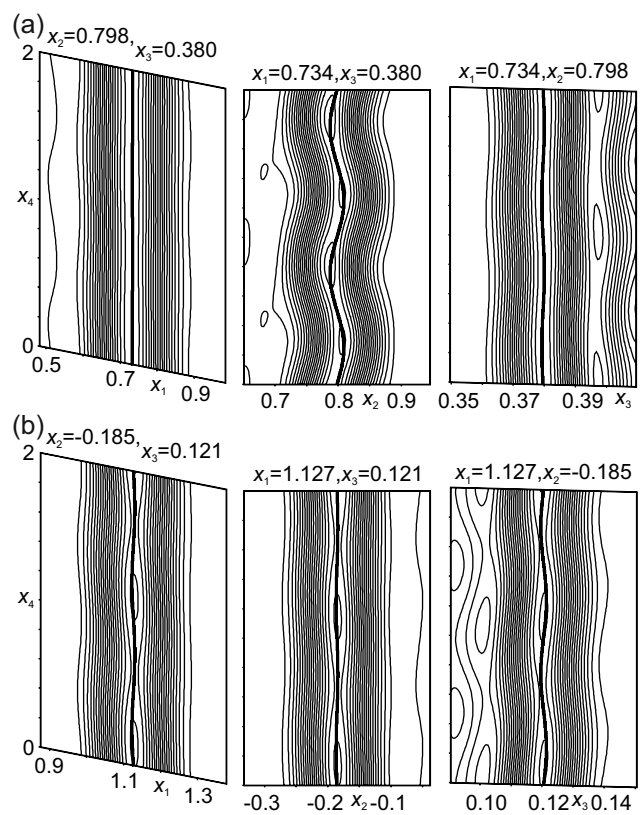}
\caption{($x_{si}$,$x_{s4}$)-sections of Fourier map centered on carbon atoms (black)
(a) C8a of molecule `A' and (b) C8b of molecule `B' respectively belonging to the phenyl ring of L-phenylalaninate moieties.
The contour line and the width of the maps are 0.5 e{\AA}$^{-3}$ and 2.5 {\AA} respectively.}
\label{f:sup_biphome_C8_fourier}
\end{center}
\end{figure}
\begin{figure}
\begin{center}
\includegraphics{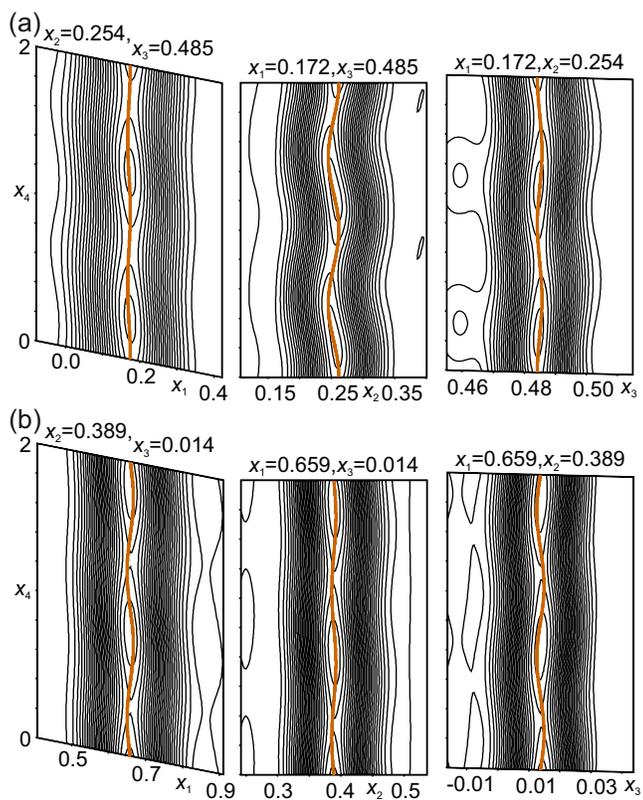}
\caption{(a) and (b) ($x_{si}$,$x_{s4}$)-sections of Fourier map centered on oxygen atoms (orange)
O2a of molecule `A' and O2b of molecule `B' respectively belonging to carboxylate groups of L-phenylalaninate moieties.
The contour line and the width of the maps are 0.5 e{\AA}$^{-3}$ and 2.5 {\AA} respectively.}
\label{f:sup_biphome_O2_fourier}
\end{center}
\end{figure}
\begin{figure}
\begin{center}
\includegraphics{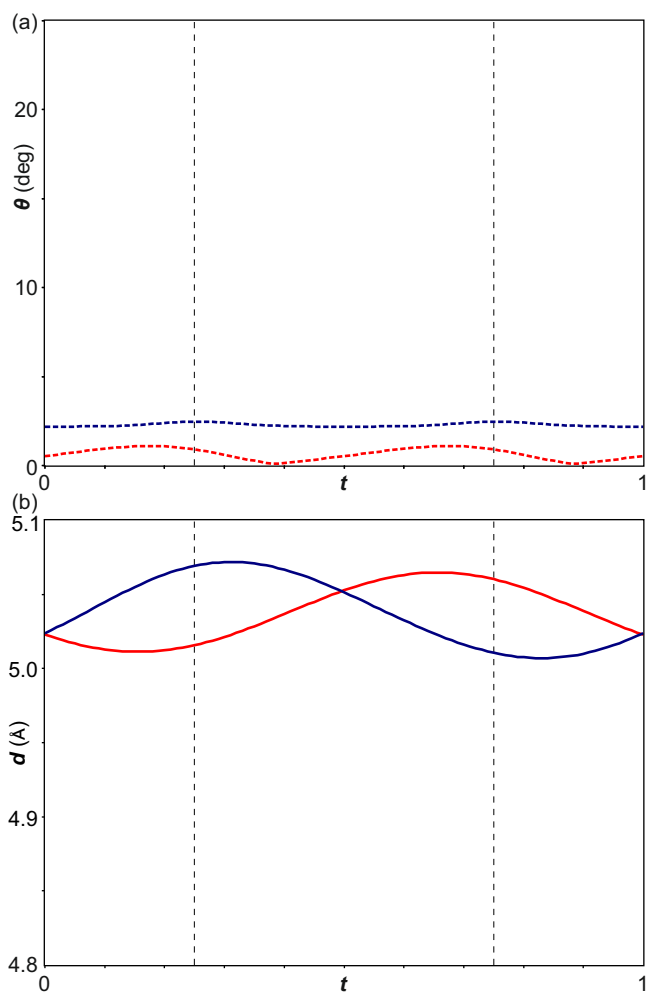}
\caption{$t$-plots of (a) angle ($\theta$) and (b) distances ($d$) describing
the tilt and intermolecular distances between phenyl rings of L-phenylalaninate moieties within stacks along $\mathbf{a}$.}
\label{f:sup_biphome_phenyl_stacking_t_plot}
\end{center}
\end{figure}
\begin{figure}
\begin{center}
\includegraphics{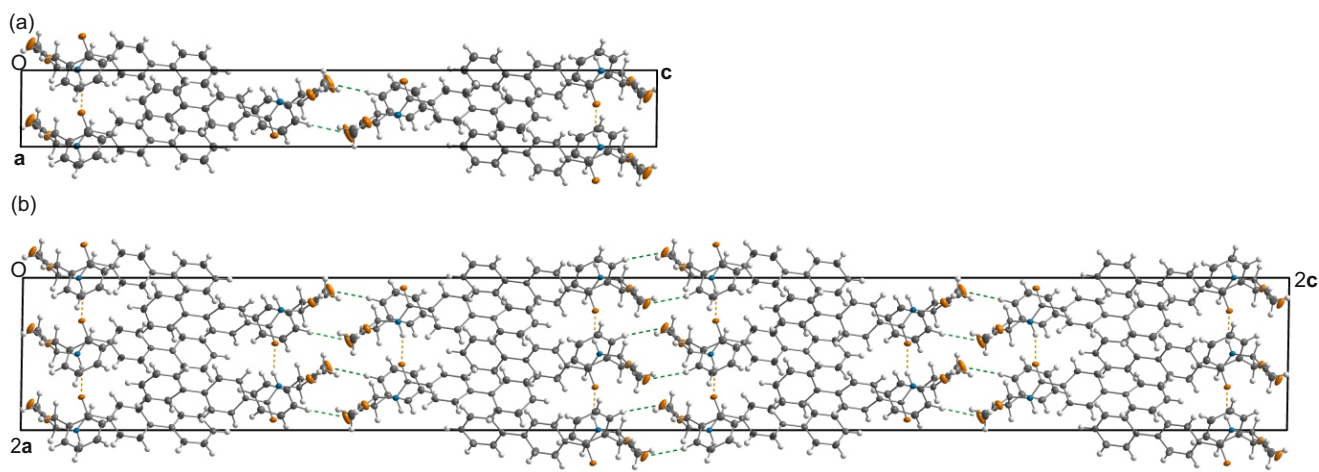}
\caption{Crystal packing of 4-biphenylcarboxy-(L)-phenylalaninate at (a) $T$ = 160 K (phase I)
and (b) $2a\times{b}\times2c$ superstructure at $T$ = 100 K (phase II).
Dashed orange lines depict linear N--H${\cdot}{\cdot}{\cdot}$O hydrogen bonds along [$\mp$100] directions,
while green dashed lines represent C--H${\cdot}{\cdot}{\cdot}$O hydrogen bond dimers.
Displacement ellipsoids are cut at 50\% probability level.
Viewing direction along [010].}
\label{f:sup_biphome_160_100_cell}
\end{center}
\end{figure}

\clearpage

\begin{table}
\begin{center}
\caption{Technical details of SCXRD measurements and number of reflections used for calculation of lattice parameters and components of modulation wave vector, $\mathbf{q}$.}
\label{t:SCXRD_measurements_reflections}
\begin{tabular}{lllll}
\hline
$T$ (K) & Number of runs & Number of images & $d_{max}$ ({\AA}) & Number of reflections \\
\hline
160     & 27             & 1345             & 0.84              & 3437                  \\
150     & 9              & 45               & 0.84              & 130                   \\
140     & 9              & 45               & 0.84              & 140                   \\
130     & 9              & 45               & 0.84              & 136                   \\
128     & 9              & 45               & 0.84              & 134                   \\
126     & 9              & 45               & 0.84              & 139                   \\
124     & 9              & 45               & 0.84              & 135                   \\
122     & 9              & 45               & 0.84              & 147                   \\
120     & 9              & 45               & 0.84              & 152                   \\
118     & 9              & 45               & 0.84              & 149                   \\
116     & 9              & 45               & 0.84              & 159                   \\
114     & 9              & 45               & 0.84              & 161                   \\
100     & 26             & 1486             & 0.84              & 3948                  \\
\hline
\end{tabular}
\end{center}
\end{table}

\begin{table}
\begin{center}
\footnotesize
\caption{Experimental and crystallographic data}
\label{t:experimental and crystallographic table}
\begin{tabular}{lll}
\hline
\multicolumn{3}{l}{Crystal data}                                                                           \\
Chemical formula                & \multicolumn{2}{c}{C$_{23}$H$_{21}$NO$_{3}$}                             \\
$M_{r}$                         & \multicolumn{2}{c}{359.42}                                               \\
%crystal size
Temperature (K)                 & 160                               & 100                                  \\
Crystal system                  & Monoclinic $b$-unique             & Monoclinic $b$-unique                \\
$a$, $b$, $c$ (\AA)             & 5.0479(2), 8.6330(4), 42.1525(15) & 5.0377(2), 8.5898(3), 42.0432(14)    \\
$\beta$ (deg)                   & 90.513(3)                         & 90.884(3)                            \\
$V$ ({\AA}$^{3}$)               & 1836.87(13)                       & 1819.11(11)                          \\
Wave vector ($\mathbf{q}$)      & --                                & $\frac{1}{2}\mathbf{a}^{*}+\frac{1}{2}\mathbf{c}^{*}$ \\
Space group                     & $P2_{1}$                          & --                                   \\
Superspace group                & --                                & $P2_{1}({\sigma}_{1}0{\sigma}_{3})0$ \\
Commensurate section            & --                                & $t_{0}$ = $\frac{1}{4}$              \\
Supercell                       & --                                & $2a{\times}b{\times}2c$              \\
Supercell space group           & --                                & $B2_{1}$                             \\
\multicolumn{3}{l}{Diffraction data}                                                                       \\
Wavelength                      & \multicolumn{2}{c}{CuK$\alpha$}                                          \\
$d$ (\AA)                       & 0.84                              & 0.84                                 \\
$\Delta{\omega}$ (deg)          & 1                                 & 1                                    \\
Absorption correction           & \multicolumn{2}{c}{multiscan}                                            \\
Criterion of observability      & \multicolumn{2}{c}{$I>3\sigma(I)$}                                       \\
\multicolumn{3}{l}{Unique reflections}                                                                     \\
all (obs/all)                   & 4219/4555                         & 5940/8898                            \\
$m$ = 0 (obs/all)               & --                                & 4150/4390                            \\
$m$ = 1 (obs/all)               & --                                & 1790/4508                            \\
$R_{int}$ (obs/all)             & 0.0200/0.0202                     & 0.0248/0.0274                        \\
$GoF$ (obs/all)                 & 1.57/1.54                         & 1.60/1.40                            \\
\multicolumn{3}{l}{$R_{F}^{obs}$/$wR_{F^{2}}^{all}$}                                                                   \\
all (obs/all)                   & 0.0393/0.0492                     & 0.0419/0.0526                        \\
$m$ = 0 (obs/all)               & --                                & 0.0368/0.0460                        \\
$m$ = 1 (obs/all)               & --                                & 0.0791/0.1191                        \\
${\Delta}{\rho}_{min}$/${\Delta}{\rho}_{max}$ (e/\AA{}$^{3}$) &     & -0.26/0.28                           \\
No. of parameters               & 494                               & 662                                  \\
H-atom treatment                & mixed                             & mixed                                \\
Twin law                        & 2 $\parallel$ $\mathbf{a}$        & 2 $\parallel$ $\mathbf{a}$           \\
Twin volumes                    & 0.9760(8)/0.0240(8)               & 0.9758(7)/0.0242(7)                  \\
\hline
\end{tabular}
\end{center}
\end{table}
\begin{table}
\small
\begin{center}
\caption{Statistical parameters
($R_{F}^{obs}$, $wR_{F^{2}}^{all}$)
of the (3+1)D incommensurately modulated (IC) and commensurately modulated (C) refinements of models
with different values of the phase $t_0$.
Number of reflections (obs/all) used in the refinements are averaged for the lowest triclinic point group symmetry:
(m=0) = 4911/5241,
(m=11) = 2031/5537.
Space group (SG) symmetries of the equivalent 3D superstructures corresponding to different C structures are given
which for the IC structure is meaningless.}
\label{t:biphome_partial_r_values_t0_sup}
\begin{tabular}{l|ll|l|ll|l}
\hline
                                                              & \multicolumn{2}{c|}{IC}  & $t_{0}=0$  & \multicolumn{2}{|c|}{$t_{0}=\frac{1}{8}$} & $t_{0}=\frac{1}{4}$  \\
\hline
SG                                                            & \multicolumn{2}{c|}{--}  & $B2_1$     & \multicolumn{2}{|c|}{$B1$}                & $B2_1$               \\
\hline
No. of parameters                                             & 811        & 649        & 649        & 811        & 649        & 649                                  \\
$GoF$(obs/all)                                                & 1.55/1.35  & 2.93/2.56  & 2.68/2.38  & 1.56/1.33  & 2.04/1.81  & 1.55/1.35                            \\
($R_{F}^{obs}$(all)                                           & 0.0428     & 0.0745     & 0.0691     & 0.0428     & 0.0566     & 0.0433                               \\
$wR_{F^{2}}^{all}$ (all)                                      & 0.0539     & 0.1035     & 0.0961     & 0.0533     & 0.0731     & 0.0547                               \\
$R_F^{obs}$($m=0$)                                            & 0.0381     & 0.0424     & 0.0401     & 0.0384     & 0.0406     & 0.0384                               \\
$wR_{F^{2}}^{all}$ ($m=0$)                                    & 0.0473     & 0.0515     & 0.0492     & 0.0475     & 0.0498     & 0.0476                               \\
$R_F^{obs}$($m=1$)                                            & 0.0777     & 0.3119     & 0.2838     & 0.0758     & 0.1750     & 0.0800                               \\
$wR_{F^{2}}^{all}$ ($m=1$)                                    & 0.1232     & 0.3174     & 0.3653     & 0.1166     & 0.2402     & 0.1280                               \\
${\Delta}{\rho}_{min}$/${\Delta}{\rho}_{max}$ (e/\AA{}$^{3}$) & -0.33/0.31 & -1.28/1.30 & -1.04/1.09 & -0.31/0.30 & -0.79/0.84 & -0.36/0.30                           \\
-ve ADPs                                                      & 4          & 4          & 1          & none       & 2          & none                                 \\
correlations $>$ 0.6                                          & 1          & 20         & 1          & 98         & 247        & 1                                    \\
\hline
\end{tabular}
\end{center}
\small
\end{table}
\begin{table}
\begin{center}
\caption{Components of the amplitude of atomic modulation functions
($|u_{x}|$, $|u_{y}|$ and $|u_{z}|$) along the three basis vectors
$\mathbf{a}$, $\mathbf{b}$ and $\mathbf{c}$ respectively
for molecules A and B.}
\label{t:amf}
\begin{tabular}{l|ll|ll|ll}
\hline
Atom & \multicolumn{2}{c|}{$|u_{x}|$ (\AA)} & \multicolumn{2}{c|}{$|u_{y}|$ (\AA)} & \multicolumn{2}{c}{$|u_{z}|$ (\AA)} \\
     & A      & B      & A      & B      & A      & B      \\
\hline
C1   & 0.0015 & 0.0086 & 0.0996 & 0.0198 & 0.0340 & 0.0092 \\
O1   & 0.0070 & 0.0050 & 0.0670 & 0.0120 & 0.0294 & 0.0071 \\
C2   & 0.0156 & 0.0166 & 0.0739 & 0.0129 & 0.0319 & 0.0198 \\
O2   & 0.0181 & 0.0433 & 0.0721 & 0.0283 & 0.0311 & 0.0483 \\
C3   & 0.0121 & 0.0081 & 0.0498 & 0.0077 & 0.0336 & 0.0134 \\
C4   & 0.0408 & 0.0045 & 0.0730 & 0.0112 & 0.0340 & 0.0172 \\
C5   & 0.0348 & 0.0070 & 0.0610 & 0.0077 & 0.0265 & 0.0244 \\
C6   & 0.0242 & 0.0010 & 0.0876 & 0.0009 & 0.0290 & 0.0311 \\
C7   & 0.0045 & 0.0066 & 0.0945 & 0.0002 & 0.0219 & 0.0286 \\
C8   & 0.0040 & 0.0196 & 0.0936 & 0.0112 & 0.0172 & 0.0378 \\
C9   & 0.0217 & 0.0388 & 0.0910 & 0.0361 & 0.0328 & 0.0563 \\
C10  & 0.0337 & 0.0136 & 0.0515 & 0.0198 & 0.0328 & 0.0357 \\
N1   & 0.0030 & 0.0247 & 0.0610 & 0.0180 & 0.0399 & 0.0387 \\
O3   & 0.0055 & 0.0045 & 0.0936 & 0.0249 & 0.0294 & 0.0345 \\
C11  & 0.0025 & 0.0035 & 0.0129 & 0.0155 & 0.0282 & 0.0189 \\
C12  & 0.0171 & 0.0141 & 0.0137 & 0.0120 & 0.0256 & 0.0210 \\
C13  & 0.1083 & 0.0171 & 0.1898 & 0.0636 & 0.0029 & 0.0433 \\
C14  & 0.1098 & 0.0146 & 0.1898 & 0.0584 & 0.0027 & 0.0391 \\
C15  & 0.0191 & 0.0257 & 0.0266 & 0.0180 & 0.0277 & 0.0202 \\
C16  & 0.1501 & 0.0821 & 0.1623 & 0.0352 & 0.0597 & 0.0130 \\
C17  & 0.1657 & 0.0539 & 0.1787 & 0.0283 & 0.0631 & 0.0029 \\
C18  & 0.0237 & 0.0247 & 0.0275 & 0.0103 & 0.0467 & 0.0219 \\
C19  & 0.1264 & 0.0640 & 0.1366 & 0.0455 & 0.0446 & 0.0042 \\
C20  & 0.1446 & 0.0423 & 0.1580 & 0.0507 & 0.0500 & 0.0105 \\
C21  & 0.0081 & 0.0332 & 0.0060 & 0.0258 & 0.0332 & 0.0399 \\
C22  & 0.0972 & 0.0463 & 0.1349 & 0.0567 & 0.0244 & 0.0483 \\
C23  & 0.0922 & 0.0348 & 0.1572 & 0.0618 & 0.0160 & 0.0416 \\
\hline
\end{tabular}
\end{center}
\end{table}
\begin{table}
\begin{center}
\caption{Equivalent value of the ADP tensors, ($U_{eq}$)
 of atoms of the biphenyl moieties at $T$ = 160 K (phase I) and $T$ = 100 K (phase II);
 and the sum of the square of the amplitudes of their atomic modulation functions along three basis vectors ($u^{2}$)
 for molecules A and B .
$u^{2}$ = $(u_{x})^{2}$ + $(u_{y})^{2}$ +$(u_{z})^{2}$.}
\label{t:adp_amf}
\begin{tabular}{l|llll}
\hline
Atom label & Molecule & $U_{eq,Phase I}$ (\AA$^{2}$) & $U_{eq,Phase II}$ (\AA$^{2}$) & $u^{2}$ (\AA$^{2}$) \\
\hline
C12        & A        & 0.0271           & 0.0193               & 0.0011              \\
           & B        & 0.0235           & 0.0182               & 0.0008              \\
C13        & A        & 0.0489           & 0.0250               & 0.0478              \\
           & B        & 0.0430           & 0.0369               & 0.0062              \\
C14        & A        & 0.0505           & 0.0267               & 0.0481              \\
           & B        & 0.0439           & 0.0371               & 0.0051              \\
C15        & A        & 0.0274           & 0.0203               & 0.0018              \\
           & B        & 0.0248           & 0.0185               & 0.0014              \\
C16        & A        & 0.0439           & 0.0256               & 0.0525              \\
           & B        & 0.0414           & 0.0334               & 0.0081              \\
C17        & A        & 0.0430           & 0.0233               & 0.0634              \\
           & B        & 0.0393           & 0.0310               & 0.0037              \\
C18        & A        & 0.0275           & 0.0193               & 0.0035              \\
           & B        & 0.0279           & 0.0202               & 0.0012              \\
C19        & A        & 0.0482           & 0.0286               & 0.0366              \\
           & B        & 0.0419           & 0.0304               & 0.0062              \\
C20        & A        & 0.0535           & 0.0314               & 0.0484              \\
           & B        & 0.0442           & 0.0311               & 0.0045              \\
C21        & A        & 0.0363           & 0.0270               & 0.0012              \\
           & B        & 0.0354           & 0.0232               & 0.0034              \\
C22        & A        & 0.0497           & 0.0311               & 0.0282              \\
           & B        & 0.0418           & 0.0293               & 0.0077              \\
C23        & A        & 0.0488           & 0.0272               & 0.0335              \\
           & B        & 0.0356           & 0.0262               & 0.0068              \\
\hline
\end{tabular}
\end{center}
\end{table}
\begin{table}
\begin{center}
\caption{Comparison of interatomic bond distances (\AA)
of molecules A and B in phase I ($T$ = 160 K) and phase II ($T$ = 100 K).}
\label{t:bond_distances}
\begin{tabular}{l|l|l|ll|ll}
\hline
            & \multicolumn{2}{c|}{phase I} & \multicolumn{4}{c}{phase II} \\
\hline
Atom groups & A    & B      & \multicolumn{2}{c|}{A} & \multicolumn{2}{c}{B} \\
\hline
            &      &                    & $t = \frac{1}{4}$ & $t = \frac{3}{4}$ & $t = \frac{1}{4}$ & $t = \frac{3}{4}$ \\
C1--O1      & 1.45 & 1.45               & 1.44          & 1.45          & 1.45          & 1.45          \\
O1--C2      & 1.31 & 1.33               & 1.32          & 1.32          & 1.33          & 1.33          \\
C2--O2      & 1.18 & 1.20               & 1.20          & 1.19          & 1.20          & 1.21          \\
C2--C3      & 1.52 & 1.53               & 1.52          & 1.52          & 1.52          & 1.52          \\
C3--C4      & 1.53 & 1.55               & 1.54          & 1.54          & 1.55          & 1.56          \\
C4--C5      & 1.51 & 1.51               & 1.50          & 1.51          & 1.51          & 1.51          \\
C5--C6      & 1.39 & 1.40               & 1.38          & 1.39          & 1.39          & 1.40          \\
C6--C7      & 1.39 & 1.38               & 1.38          & 1.38          & 1.38          & 1.38          \\
C7--C8      & 1.37 & 1.38               & 1.39          & 1.40          & 1.38          & 1.38          \\
C8--C9      & 1.38 & 1.39               & 1.38          & 1.38          & 1.40          & 1.39          \\
C9--C10     & 1.39 & 1.40               & 1.39          & 1.39          & 1.38          & 1.38          \\
C10--C5     & 1.39 & 1.38               & 1.40          & 1.39          & 1.39          & 1.39          \\
C3--N1      & 1.45 & 1.45               & 1.45          & 1.46          & 1.45          & 1.45          \\
N1--C11     & 1.34 & 1.33               & 1.33          & 1.33          & 1.32          & 1.33          \\
C11--O3     & 1.23 & 1.23               & 1.24          & 1.24          & 1.24          & 1.23          \\
C11--C12    & 1.50 & 1.50               & 1.50          & 1.50          & 1.50          & 1.51          \\
C12--C13    & 1.37 & 1.37               & 1.39          & 1.38          & 1.37          & 1.37          \\
C13--C14    & 1.39 & 1.39               & 1.39          & 1.38          & 1.39          & 1.39          \\
C14--C15    & 1.38 & 1.38               & 1.40          & 1.39          & 1.39          & 1.39          \\
C15--C16    & 1.38 & 1.37               & 1.40          & 1.38          & 1.38          & 1.38          \\
C16--C17    & 1.39 & 1.39               & 1.39          & 1.39          & 1.38          & 1.39          \\
C17--C12    & 1.37 & 1.38               & 1.39          & 1.38          & 1.39          & 1.38          \\
C15--C18    & 1.49 & 1.50               & 1.48          & 1.49          & 1.50          & 1.50          \\
C18--C19    & 1.39 & 1.40               & 1.39          & 1.40          & 1.39          & 1.39          \\
C19--C20    & 1.39 & 1.38               & 1.38          & 1.38          & 1.39          & 1.38          \\
C20--C21    & 1.36 & 1.38               & 1.39          & 1.39          & 1.37          & 1.38          \\
C21--C22    & 1.36 & 1.37               & 1.38          & 1.37          & 1.37          & 1.37          \\
C22--C23    & 1.39 & 1.38               & 1.38          & 1.39          & 1.37          & 1.39          \\
C23--C18    & 1.38 & 1.39               & 1.39          & 1.40          & 1.40          & 1.39          \\
\hline
\end{tabular}
\end{center}
\end{table}

\begin{table}
\begin{center}
\caption{Comparison of lattice parameters and residual values from Le baile fit
of the PXRD pattern based on two unit cells.
Lattice parameters obtained from SCXRD data has been given as reference.}
\label{t:lat_par_compare}
\begin{tabular}{llll}
\hline
                  & SCXRD       & \multicolumn{2}{c}{PXRD}        \\
                  &             & Cell 1        & Cell 2          \\
$a$ (\AA)         & 5.0646(2)   & 14.5357(12)   & 13.7041(10)     \\
$b$ (\AA)         & 8.7483(3)   & 8.6153(6)     & 13.0856(9)      \\
$c$ (\AA)         & 42.4157(15) & 16.5541(12)   & 11.4692(8)      \\
$\alpha$ (deg)    & 90          & 108.248(5)    & 89.337(7)       \\
$\beta$ (deg)     & 90          & 103.902(4)    & 99.307(5)       \\
$\gamma$ (deg)    & 90          & 80.233(5)     & 89.436(7)       \\
$V$ ({\AA}$^{3}$) & 1879.27(12) & 1901.1(3)     & 2029.4(3)       \\
$GoF$             & --          & 3.51          & 3.53            \\
$R_{p}/wR_{p}$    & --          & 0.0422/0.0642 & 0.0416/0.0646   \\
\hline
\end{tabular}
\end{center}
\end{table}

\clearpage % force a pagebreak and flush all deferred `table` and `figure` environments

%\bibliography{biphome}

\providecommand{\noopsort}[1]{}\providecommand{\singleletter}[1]{#1}%